\documentclass[12pt]{article}
\usepackage{yfonts}
\usepackage{amsmath}
\usepackage{amssymb}
\usepackage{amstext}
\usepackage{amscd}
\usepackage{amsfonts}
\usepackage{color}

\newcommand{\be}{\begin{equation}}
\newcommand{\nd}{\noindent}
\newcommand{\ee}{\end{equation}}
\newcommand{\ben}{\begin{eqnarray}}
\newcommand{\een}{\end{eqnarray}}

\title{{\small{\bf On the nature of the Tsallis-Fourier Transform}}}

\author{\small{A. Plastino and M. C. Rocca} \\ \small{
La Plata Physics Institute - CCT-Conicet-Exact Sciences Faculty}
\\
\small {Universidad Nacional (UNLP), C.C. 727 (1900) La Plata,
Argentina}}

\begin{document}

\maketitle
\begin{abstract}

\nd  {\small By recourse to tempered ultradistributions, we show
here that the effect of a q-Fourier transform (qFT) is to map {\it
equivalence classes} of functions into other classes in a
one-to-one fashion. This suggests that Tsallis' q-statistics may
revolve around equivalence classes of distributions and not on
individual ones, as orthodox statistics does. We  solve here the
qFT's non-invertibility issue, but discover a problem that remains
open.}

\nd \small{KEYWORDS: q-Fourier transform, tempered
ultradistributions.}

\end{abstract}

\newpage

\renewcommand{\theequation}{\arabic{section}.\arabic{equation}}

\section{Introduction}

\nd    Non-extensive statistical mechanics (NEXT)
\cite{[1],[2],AP}, a well known generalization of the
Boltzmann-Gibbs (BG) one, is used in many scientific and
technological endeavors. NEXT central concept is that of a
nonadditive (though extensive \cite{[3]}) entropic information
measure characterized by the real index q (with q = 1 recovering
the standard BG entropy). Applications include  cold atoms in
dissipative optical lattices \cite{[4]}, dusty plasmas \cite{[5]},
trapped ions \cite{[6]}, spin glasses \cite{[7]}, turbulence in
the heliosphere \cite{[8]}, self-organized criticality \cite{[9]},
high-energy experiments at LHC/CMS/CERN \cite{[10]} and
RHIC/PHENIX/Brookhaven \cite{[11]}, low-dimensional dissipative
maps \cite{[12]}, finance \cite{[13]}, galaxies \cite{AP1},
 and Fokker-Planck equation's studies \cite{AP2}, EEG's \cite{martin}, complex signals \cite{martin1},
 Vlasov-Poisson equations \cite{arpalp}, etc.\vskip 3mm

\noindent \fbox{\parbox{5.4in}{So called q-Fourier transforms
(qFT) were developed by Tsallis et al. \cite{umarov}.
They constitute  a central piece in the Tsallis' q-machinery.
However (see \cite{PR} and references therein), \color{blue}  qFT
seems not to be an invertible transformation.}} \normalcolor
\vskip 3mm

\nd The imaginary q-exponential function, a protagonist of Tsallis'
statistics, is defined as

\be    e_q(ix) = \left[1 + i (1-q)x\right]^{\frac{1}{1-q}}, \ee with

\be e_q(ix) \rightarrow \exp{(ix)} \,\,\,\,{\rm whenever} \,\,\,\,q
\rightarrow 1.\ee

\noindent \fbox{\parbox{5.4in}{ However, the function \be
\label{family} e_q^b(ix)=  \left[1 +i
(1-q)x\right]^{\frac{1}{1-q}+b}, \ee with $-\infty<b<0$ , $b$ a
real number also fulfills \be e_q^b(ix) \rightarrow \exp{(ix)}
\,\,\,\,{\rm whenever} \,\,\,\,q \rightarrow 1.\ee Accordingly,
there is a {\it class} of functions $ e_q^b(ix)$, labelled by $b$,
that tend to the ordinary exponential in the limit $q \rightarrow
1$. This fact profoundly affects the workings of the q-Fourier
Transform.}} \vskip 3mm \nd  The same happens with the q-logarithm
defined as \be \ln_q(x)=\frac {x^{1-q}-1} {1-q}\rightarrow\ln
(x)\; {\rm whenever}\; q\rightarrow 1 \ee The function \be
\ln^c_q(x)=\frac {x^{\frac {1-q} {1+c(1-q)}}-1} {1-q}
\rightarrow\ln (x)\; {\rm whenever}\; q\rightarrow 1 \ee \nd where
$-\infty<c<0$ is a real number. Moreover \be
\ln^b_q[e_q^b(ix)]=ix. \ee

\nd Recall that Schwartz space $\mathcal{S}$ is the
space of functions all of whose derivatives are rapidly decreasing
\cite{tp1}. This space has the important property that the Fourier
transform is an automorphism on this space. This property enables
one, by duality, to define the Fourier transform for elements in
the dual space of $\mathcal{S}$, called the space of tempered
distributions \cite{tp1}. \vskip 3mm

\nd \color{blue} In the present communication we reconcile the
Tsallis et al. developments \cite{umarov} with the
non-invertibility issue \normalcolor and show, by recourse to
tempered ultradistributions (a generalization and extension to the
complex plane of Schwartz' tempered distributions), that the qFT
does indeed map, in a one-to-one fashion, {\it classes} of
functions into other classes, not isolated functional instances.
Thus, such issue can be resolved by appealing to a higher order of
mathematical perspective.  Section 2 recapitulates the findings of
our Ref. \cite{PR}, related to an extension of the Tsallis et al.
environment \cite{tt1}, that becomes just a particular case (real
line) of a more encompassing theory (complex plane) \cite{PR}.
Section 3, the core of our presentation, specializes the
developments of \cite{PR} to an important situation. We analyze
there the particular instantiation of the theory that leads to the
scenario investigated by both Tsallis et al. \cite{umarov}. The
results thus obtained, our main contribution here, are illustrated
in Section 5 by an important example. Finally, the mathematical
problem that remains open is discussed in Section 5. Conclusions
are drawn in Section 6.

\setcounter{equation}{0}

\section{Reviewing an  alternative qFT definition}

\nd  In Ref. \cite{PR} (see also \cite{tt1}) we introduced an
alternative qFT-definition, that, by generalizing and extending
the original one,  overcomes the non-invertibility problem
afflicting the one of \cite{umarov}. We briefly review that
alternative version in this Section. Our protagonists in such
endeavor were tempered ultradistributions \cite{tt1,tp1,tp2,tp6},
that constitute a generalization of the distributions-set for
which the test functions are members of a Schwartz space
$\mathcal{S}$, a function-space in which its members possess
derivatives that are rapidly decreasing. $\mathcal{S}$ exhibits a
notable  property: {\it the Fourier transform is an automorphism
on}  $\mathcal{S}$, a property that allows, by duality, to define
the Fourier transform for elements in the dual space of
$\mathcal{S}$. {\it This dual is the space of tempered
distributions} \cite{tp1}. \vskip 4mm

\nd In physics it is not uncommon to face functions that grow
exponentially in space or time. In such circumstances Schwartz'
space of tempered distributions is too restrictive. One needs {\it
ultradistributions} to generate appropriate answers
\cite{bollini}. They are  continuous linear functionals defined on
the space of entire functions rapidly decreasing on straight lines
parallel to the real axis \cite{bollini}.  Now, following
\cite{tt1}, we appeal to
 the Heaviside step function $H$

  \ben \label{p1}
H(x)=
\begin{cases}
1 \,\,\,\,\, for \,\,\,\,\, x \ge 0 \\
0 \,\,\,\,\, for \,\,\,\,\, x < 0.
\end{cases}
 \een
\nd \fbox{\parbox{5.1in}{These step functions allow us to properly
define the q-Fourier transform  in a different fashion as that of
Tsallis et al. \cite{umarov} by recourse  \cite{tt1} to the space
$\boldsymbol{{\cal U}}$ of tempered ultradistributions [see Ref.
\cite{tp2}]}} \vskip 3mm

 \nd As stated above, a tempered ultradistribution is
a continuous linear functional defined on the space
$\boldsymbol{{\cal H}_1}$ of entire functions rapidly decreasing
on straight lines parallel to the real axis.
Let $\Omega$ be the space of functions of the real variable $x$
that are parametrized by a real parameter $q$:
\begin{equation}
\label{ep1.1}
\Omega=\{f_q(x)/f_q(x)\in{\Omega}^+\cap{\Omega}^-\},
\end{equation}
where
\[{\Omega}^+=\left\{f_q(x)/f_q\mid_{{\mathbb{R}}^+}(x)
\{1+i(1-q^{'})kx[f_q\mid_{{\mathbb{R}}^+}(x)
]^{(q^{'}-1)}\}^{\frac {1} {1-q^{'}}}\in
{\cal L}^1[\mathbb{R}^+];\right.\]
\[f_q(x)\geq 0; \left|f_q(x)\right|\leq |x|^pg(x)e^{ax};p,a\in\mathbb{R}^+;
k\in\mathbb{C};\Im(k)\geq 0\]
\begin{equation}
\label{ep1.2} \left.1\leq q,q^{'}<2\right\},
\end{equation}
and
\[{\Omega}^-=\left\{f_q(x)/f_q\mid_{{\mathbb{R}}^-}(x)
\{1+i(1-q^{'})kx[f_q\mid_{{\mathbb{R}}^-}(x)
]^{(q^{'}-1)}\}^{\frac {1} {1-q^{'}}}\in
{\cal L}^1[\mathbb{R}^-];\right.\]
\[f_q(x)\geq 0; \left|f_q(x)\right|\leq |x|^pg(x)e^{ax};p,a\in\mathbb{R}^+;
k\in\mathbb{C};\Im(k)\leq 0\]
\begin{equation}
\label{ep1.3} \left.1\leq q,q^{'}<2\right\},
\end{equation}
Here ${\cal L}^1$ is the space of functions integrable
in Lebesgue sense,
$g(x)$  is bounded, continuous, and positive-definite,
$f_q\mid_{{\mathbb{R}^+}}(x)$ is the restriction of $f_q(x)$
to ${{\mathbb{R}^+}}$ and
$f_q\mid_{{\mathbb{R}^-}}(x)$ is the restriction of $f_q(x)$
to ${{\mathbb{R}^-}}$.

\vskip 3mm

\noindent \fbox{\parbox{5.4in}{Our q-Fourier transform is now
defined ($\Im(k)$ is the imaginary part of $k$) as
\begin{equation}
\label{ep1.4} F:\Omega\longrightarrow\boldsymbol{{\cal
U}}\,\,\,\,(the\,\, space\,\, of\,\, tempered\,\,
ultradistributions),
\end{equation}
where:
\begin{equation}
\label{ep1.5} F(f_q)(k,q^{'}) \equiv F(k,q^{'},q),
\end{equation}
with:
\[F(k,q^{'},q)=[H(q^{'}-1)-H(q^{'}-2)]\times \]
\[\left\{H[\Im(k)]\int\limits_0^{\infty} f_q(x)
\{1+i(1-q^{'})kx[f_q(x)]^{(q^{'}-1)}\}^{\frac {1}
{1-q^{'}}},
\;dx -\right.\]
\begin{equation}
\label{ep1.6} \left. H[-\Im(k)]\int\limits_{-\infty}^0 f_q(x)
\{1+i(1-q^{'})kx[f_q(x)]^{(q^{'}-1)}\}^{\frac {1} {1-q^{'}}} \;dx\right\}.
\end{equation}}}
The inverse transformation is (\cite{PR},\cite{tt1}):
\begin{equation}
\label{ep1.7}
f_q(x)=\frac {1}
{2\pi}\oint\limits_{\Gamma}\left[\lim_{\epsilon\rightarrow 0^+}
\int\limits_1^2 F(k,q^{'},q)\delta (q^{'}-1-\epsilon)\;dq^{'}\right]
e^{-ikx}\;dk.
\end{equation}
{\it As has been proved in \cite{PR},\cite{tt1}, $F$ is one to one
from} $\Omega$ to $\boldsymbol{{\cal U}}$ \vskip 3mm \nd On the
real axis:
\[F(k,q^{'})=[H(q^{'}-1)-H(q^{'}-2)]\times \]
\begin{equation}
\label{ep1.8} \int\limits_{-\infty}^{\infty} f_q(x)
\{1+i(1-q^{'})kx[f_q(x)]^{(q^{'}-1)}\}^{\frac {1} {1-q^{'}}} \;dx,
\end{equation}
for the real transform, and
\begin{equation}
\label{ep1.9} f_q(x)=\frac {1}
{2\pi}\int\limits_{-\infty}^{\infty}\left[\lim_{\epsilon\rightarrow
0^+} \int\limits_1^2 F(k,q^{'},q)\delta (q^{'}-1-\epsilon)\;dq{'}\right]
e^{-ikx}\;dk,
\end{equation}
for its inverse.

\setcounter{equation}{0}

\section{q-FT in the limit $q'
\rightarrow q$}

\nd {\bf Our main result is to be presented now}, by consideration
of the  limit $q' \rightarrow q$. \vskip 3mm

\noindent \fbox{\parbox{5.4in}{This leads to the (restricted)
scenario in which the Tsallis et al. {\bf non-invertibility issue}
raises its head.}}

\vskip 3mm  \nd Define the restricted (i.e., to the $q'=q$
situation) transform, i.e., the Tsallis et al. one,
\begin{equation}
\label{ep1.10}
F_T:\Omega\longrightarrow\boldsymbol{{\cal U}}
\end{equation}
as
\begin{equation}
\label{ep1.11}
F_T(f_q)(k)=\lim_{q^{'}\rightarrow q}F(f_q)(k,q^{'})=
F(f_q)(k,q^{'}){\mid}_{q^{'}=q}
\end{equation}
Thus, according to (\ref{ep1.6}),
\[F_T(k,q)=[H(q-1)-H(q-2)]\times \]
\[\left\{H[\Im(k)]\int\limits_0^{\infty} f_q(x)
\{1+i(1-q)kx[f_q(x)]^{(q-1)}\}^{\frac {1}
{1-q}},
\;dx -\right.\]
\begin{equation}
\label{ep1.12} \left. H[-\Im(k)]\int\limits_{-\infty}^0 f_q(x)
\{1+i(1-q)kx[f_q(x)]^{(q-1)}\}^{\frac {1} {1-q}} \;dx\right\}
\end{equation}

\nd \fbox{\parbox{4.7in}{It is seen in \cite{PR} that $F_T$ is NOT
one to one from $\Omega$ to $\boldsymbol{{\cal U}}$.}}

\vskip 3mm  \nd The problem is best understood is we introduce a
particularly important set ${\Lambda}_{f_q}$, crucial for our
considerations.  Let ${\Lambda}_{f_q}$ be  given by
\begin{equation}
\label{ep1.13} {\Lambda}_{f_q}=\left\{g_q\in\Omega/
F_T(g_q)(k)=F_T(f_q)(k)\right\},
\end{equation}
and
\begin{equation}
\label{ep1.14}
\Lambda=\left\{{\Lambda}_{f_q}/f_q\in\Omega\right\}.
\end{equation}
We define the {\it equivalence relation}
\begin{equation}
\label{ep1.15}
g_q(x)\sim f_q(x)\Longleftrightarrow g_q\in\Lambda_{f_q}
\end{equation}
and, subsequently, {\bf the special the version of the
Tsallis et al. q-Fourier transform} $F_{UTS}$
\cite{umarov}
\begin{equation}
\label{ep1.16}
F_{UTS}:\Lambda\longrightarrow\boldsymbol{{\cal U}}
\end{equation}
as
\begin{equation}
\label{ep1.17} F_{UTS}(\Lambda_{f_q})(k)=F_T(f_q)(k).
\end{equation}

\nd \fbox{\parbox{5.1in}{ We see now that $F_{UTS}$ is an
application from {\it equivalence classes into equivalence
classes} and, as a consequence, one to one from $\Lambda$ into
$\boldsymbol{{\cal U}}$!}} \vskip 3mm

\nd We realize now that the Tsallis et al. qFT is actually a
(one-to-one) set-to-set transformation, which solves the
non-invertibility issue  that occupies our attention in this work.

\setcounter{equation}{0}

\section{Illustration}
\nd Illustrating  our theory we reconsider an important example.
 Focus attention of the so called Hilhorst function (see pertinent details in
\cite{PR} and references therein)

\begin{equation} \label{ep1.18}
f_q(x)=
\begin{cases}
\left(\frac {\lambda} {x}\right)^{\beta}\;;\; x\in[a,b]\;;\; 0<a<b\;;\;\lambda>0 \\
0\;;\;x\; \rm{outside}\; [a,b],
\end{cases}
\end{equation}
with
\begin{equation}
\label{ep1.19} \lambda=\left[\left(\frac {q-1} {2-q}\right)
\left(a^{\frac {q-2} {q-1}}-b^{\frac {q-2} {q-1}}
\right)\right]^{1-q}\;\;\;\beta=\frac {1} {q-1}.
\end{equation}

\nd In Ref.  \cite{tt1} we evaluated the q-Fourier transform on
this function and  obtained
\[F(k,q^{'},q)=[H(q^{'}-1)-H(q^{'}-2)]H[\Im(k)]\times\]
\[\left\{\left\{H(q^{'}-1)-H\left[q-\left(1+\frac {1} {\beta}\right)\right]\right\}\right.\times\]
\[\frac {(q^{'}-1){\lambda}^{\beta}} {(2-q^{'})
[(1-q^{'})ik{\lambda}^{\beta}]^{\frac {1} {q^{'}-1}}}\times \]
\[\left\{a^{\frac {q^{'}-2} {q^{'}-1}}F\left(\frac {1} {q^{'}-1},
\frac {2-q^{'}} {(q^{'}-1)[1-\beta(q^{'}-1)]},
\frac {1} {q^{'}-1} + \frac {\beta(2-q^{'})} {1-\beta(q^{'}-1)};\right.\right.\]
\[\left.\frac {1} {(q^{'}-1)ik{\lambda}^{\beta(q^{'}-1)}a^{1-\beta(q^{'}-1)}}\right)-\]
\[ b^{\frac {q^{'}-2} {q^{'}-1}}F\left(\frac {1} {q^{'}-1},
\frac {2-q^{'}} {(q^{'}-1)[1-\beta(q^{'}-1)]},
\frac {1} {q^{'}-1} + \frac {\beta(2-q^{'})} {1-\beta(q^{'}-1)};\right.\]
\[\left.\left.\frac {1} {(q^{'}-1)ik{\lambda}^{\beta(q^{'}-1)}b^{1-\beta(q^{'}-1)}}\right)\right\}+\]
\[\left\{H\left[q^{'}-\left(1+\frac {1} {\beta}\right)\right]-H(q^{'}-2)\right\}
\frac {{\lambda}^{\beta}} {\beta-1}\times\]
\[\left\{a^{1-\beta}F\left(\frac {1} {q^{'}-1},\frac {\beta-1} {\beta(q^{'}-1)-1},
\frac {\beta q^{'}-2} {\beta(q^{'}-1)-1};\right.\right.\]
\[\left.(q^{'}-1)ik{\lambda}^{\beta(q^{'}-1)}a^{1-\beta(q^{'}-1)}\right)-\]
\[b^{1-\beta}F\left(\frac {1} {q^{'}-1},\frac {\beta-1} {\beta(q^{'}-1)-1},
\frac {\beta q^{'}-2} {\beta(q^{'}-1)-1};\right.\]
\begin{equation}
\label{ep1.20}
\left.\left.\left.(q^{'}-1)ik{\lambda}^{\beta(q^{'}-1)}b^{1-\beta(q^{'}-1)}\right)\right\}\right\}.
\end{equation}
Taking $q^{'}=q$ in (\ref{ep1.20}) we have for $F_{UTS}$:
\begin{equation}
\label{ep1.21}
F_{UTS}(\Lambda_{f_q})(k)=H[\Im(k)]\left[H(q-1)-H(q-2)\right]
\left[1+(1-q)ik\lambda\right]^{\frac {1} {1-q}},
\end{equation}
and, on the real axis,
\[F_{UTS}(\Lambda_{f_q})(k)=\left[H(q-1)-H(q-2)\right]
\left[1+(1-q)i(k+i0)\lambda\right]^{\frac {1} {1-q}}=\]
\begin{equation}
\label{ep1.22} \left[H(q-1)-H(q-2)\right]
\left[1+(1-q)ik\lambda\right]^{\frac {1} {1-q}}.
\end{equation}

\vskip 2mm \nd \fbox{\parbox{5.1in}{ Now, from (\ref{ep1.21}) and
(\ref{ep1.22}) we see that the Tsallis et al. q-Fourier
transform is one to one. But it is a transformation from $\Lambda$
into $\boldsymbol{{\cal U}}$, a class-to-class one.}}

\vskip 2mm  \nd This fact  reconciles the viewpoints Tsallis et al.
those of \cite{PR} and this is achieved via a rigorous definition
of the qFT and of its domain and image.

\nd As a second example we consider

\[f(x)=H(x).\]
In this case, the equivalence class is made up of just one
function:
\begin{equation}
\label{epu.12} F_{UTS}(k,q)=[H(q-1)-H(q-2)]H[\Im(k)]\int\limits_0^\infty
\left[1+(1-q)ikx\right]^{\frac{ 1} {1-q}}\;dx.
\end{equation}
Evaluating the integral we have:
\begin{equation}
\label{epu.13} F_{UTS}(k,q)=[H(q-1)-H(q-2)]H[\Im(k)]\frac {\Gamma\left(\frac {2-q}
{q-1}\right)} {\Gamma\left(\frac {1}
{q-1}\right)}\left[(1-q)ik\right]^{-1},
\end{equation}
and, finally,
\begin{equation}
\label{epu.14} F_{UTS}(k,q)=[H(q-1)-H(q-2)]\frac {i} {2-q} \frac {H[\Im(k)]} {k}.
\end{equation}

\setcounter{equation}{0}

\section{An open problem}

\nd  Let $\hat{\sim}$ be the equivalence relation defined by:
\begin{equation}
\label{ep5.1} g_q\hat{\sim} f_q \Longleftrightarrow
\lim_{q\rightarrow 1} g_q= \lim_{q\rightarrow 1}f_q,
\end{equation}
where $f_q,g_q\in\Omega$. Let
\begin{equation}
\label{ep5.2} \varXi_{f_q}=\{g_q\in\Omega/g_q\hat{\sim} f_q\},
\end{equation}
and let
\begin{equation}
\label{ep5.3} \varXi=\{\varXi_{f_q}/f_q\in\Omega\}.
\end{equation}
If $f_q=f_{1q}+ (q-1)f_{2q}$ , $g_q=g_{1q}+ (q-1)g_{2q}$ and
$\lim_{q\rightarrow 1}f_{1q}=\lim_{q\rightarrow 1}g_{1q}$ then
$f_q,g_q\in\varXi_{f_q}$. However, since $f_{2q},g_{2q}\in\Omega$
are different in general, one does not have $f_q\sim g_q$.
As a simple example of this take $f_{1q}(x)=f_{2q}(x)=
x^{q-1}$ and $g_{1q}=g_{2q}=x^{2(q-1)}$.
Accordingly,  $\Lambda_{f_q}\subseteq\varXi_{f_q}$.

\nd One would, of course, be  interested in finding out which are
the mathematical properties of the functions $f_q$ and $g_q$ that
generate the belonging  $g_q\in\Lambda_{f_q}$.
In other words, what
are the mathematical properties that the functions
$f_q$ and $g_q$
need to have so that $F_T(f_q)=F_T(g_q)$. Note that one has
\begin{equation}
\label{ep5.4} F_T(f_q)=F_T(g_q)\Longrightarrow
F(f)=F(g)\Longrightarrow f=g.
\end{equation}
That is $\lim_{q\rightarrow 1}f_q=\lim_{q\rightarrow 1}g_q.$ This
is a problem that we were unable to solve and that we would like
to be considered by the mathematical  community

\section*{Conclusions}

\nd We have shown here an important original result:  the
q-generalization advanced by Tsallis et al. in \cite{umarov} is to
be properly regarded as {\it a transformation between classes of
equivalence} and thus one-to-one, {\bf a finding of this paper
that solves the qFT's non-invertibility issue} \cite{PR}.

\nd Our present findings may indicate that Tsallis' q-statistics
revolves around equivalence classes of distributions and not on
individual ones, as orthodox statistics does. In Section 5 we have
seen, however, that an open problem remains that should be
addressed in the future.

\setcounter{equation}{0}

\end{document}